# FPGA-based Implementation of a New Data Frame Correction System for Merging Units

Mohammad Hashemi, and Bijan Alizadeh, *Senior Member, IEEE*

*Abstract*— With today's increasing demand for digital devices in Substation Automation Systems (SAS) based on the IEC61850 standard, the measured data error due to the synchronization problem should be considered as a significant problem in digitalized SAS. Although time tagging and mathematical methods have been proposed to alleviate this problem, they require a massive amount of calculations and elaborations. To develop a solution for both problems of the data error and the massive computation, in this paper, we propose a data frame correction (DFC) system with a new method of data shift computation as a data correction method implemented as a hardware accelerator on FPGA. Compared to the state-of-the-art DFC systems, the results show that the proposed DFC system can achieve data correction with up to 99.6% fewer hardware resources utilization and fulfills 9× calculation speed while maintaining IEC61850 required accuracy in 2.1ms.

## I. Introduction

WITH, the proliferation of the IEC61850 standard [1], digital-based substation automation systems (SAS) have earned special attention in electrical engineering. The intelligent electronic devices (IEDs) process data and communicate with each other using digitalized data but substation automation systems inputs are of the waveform; therefore, there is a need for a unit, Merging Unit (MU) [2], equipped with a network interface to provide digitalized data and delivers it to the IEDs [3]. By considering at least six MUs in each SAS [4], each pair of MUs is responsible for sampling the Voltage and Current data from each power transmission line (TL). All MUs must be synchronized and have the same time tagging system to avoid data shift errors [5]. The latency of the measurement systems and inaccuracy of the MU clock cause measured data shift errors, resulting in three significant problems. The first problem is that the data shift error directly affects wasting energy [6], up to 540MW as mentioned in [6], because the unreliability of the measured data causes error in energy monitoring. The second issue is that if an event occurs in the TL, the system monitoring units anticipate the location of the fault based on the input. So, the unreliability of the measured data and data shift error causes the fault location anticipation error. This results in the maintenance cost [7], every 1-microsecond causes fault location offset of 500 meters in the fault location calculation [8]. The third and the most crucial problem is that many events in the bay level of the SAS occur in the order of nanoseconds (ns) [9]. A shift in the data sampling may cause event detection failure. To avoid this problem, each SAS should have a mechanism that guarantees 100% accuracy of the data sampling. At present, IRIGB [9], the Master-slave synchronization method [10], and IEEE1588 [11] are the most potent mechanisms to maintain the accuracy of the data sampling in SAS. However, even the most potent mechanism has an error, especially in harsh environments where the measurement delay is about one millisecond (ms), which is inevitable [12]. Hence, a data frame correction (DFC) system needs to minimize data shift errors before sending them to IEDs. It should be noted that, in the rest of the paper, the measurement delay is the summation of measurement devices intrinsic delays and MU sampling delay.

Most of the available DFC systems find data shift by comparing the amplitude of measured data and a reference signal. Authors of [4] choose the data amplitude of one of MUs in SAS as the reference signal and calculate all other MUs' data shift errors compared to the chosen MU data amplitude. Its drawback is that the chosen MU's data amplitude, which is considered as the reference signal, could have data shift itself, and this data shift could cause an error in the data frame correction. In another approach, the authors of [13] estimate the reference signal based on the equivalent circuit of the TLs. Then, data shift is obtained by comparing the amplitude of measured data and estimated reference signal amplitude. As each TL has its unique equivalent circuit, the drawback of [13] is that a new DFC system should be designed for each TL. Authors of [14] estimate the TLs signal using the Kalman filter model [15] and found the difference between the measured data and the estimated signal. As the Kalman filter model is a recursive method of estimation based on prior TL signal values and the accuracy of the DFC system in [14] is not 100%, the error that remained in previous corrected frames causes an additive error in later frames. In another approach, the authors of [9] increase the sampling ratio of measured data and use Kaisor's window [16] to extract measured data features and compare them with reference features sent by Supervisory Control and Data Acquisition System (SCADA). If there is any difference between extracted and reference features, the data shift error will be removed using the estimated time sampling (ETS) method as a curve fitting technique. As the ETS method requires a massive computation, the drawback of [9] is that it needs a colossal hardware resource utilization to be implemented.

The downside of the above-mentioned amplitude-based DFC systems is their dependency on a reference signal. Obtaining this reference signal requires a huge amount of calculations. Moreover, the reference signal is an approximation of the actual TL signal; therefore, it acts as an origin of error. To alleviate these problems, in this work, we remove the data shift error based on the measured data phase without the requirement of a reference signal. The MU takes samples (measured data) from amplitude of the TL signal. As the MU takes these samples with a delay (measurement delay), all measured data have a data shift error, and this error is a function of measurement delay [17]. Hence, the idea behind this work is to calculate and remove data shift error by finding the measurement delay and re-framing the measured data. This way, we present a DFC system that finds the measurement delay by calculating MU data sampling delay and uses interpolation techniques to re-frame measured data. Note that the response time of the DFC system must be less than 10 microseconds (us) [18] which is challenging for software-based implementations on CPU [4]. Thus, we benefit from the field-programmable gate arrays (FPGA) ability of parallel computation and implement the proposed DFC system on FPGA. Hence, the main contributions of this paper are as follows:

- Proposing a new DFC system that removes the data shift of measured data frame based on measured delay phase without reference signal requirement.
- Presenting a novel architecture for hardware implementation of the proposed DFC system with less resource utilization, up to 99.6%, and faster response time, up to 9×, compared to the state-of-the-art DFC systems.

The rest of the paper is organized as follows. Section II describes the preliminaries. The proposed DFC system and related FPGA-based architecture are explained in Sections III and IV, respectively. Experimental results are reported in Section V, and Section VI concludes the paper.

## II. PRELIMINARIES

The electricity is generated in power plants and transmitted through TL as an alternating current (AC) signal (TL signal) [17]. As the power system frequency is 50Hz [17], the TL signal is generated as an AC periodic signal with a period of $\frac{1s}{50} = 20ms$. The amplitude of the TL signal in each period of TL signal must be monitored and evaluated in SAS [2]. The IEC61850 standard part-9-2 asserted that a set of 256 samples (defined as measured data frame) of each period of TL signal are adequate for monitoring and evaluation process in SAS [2]. Hence, the MU takes 256 samples (256 measured data) of each TL signal period [2]. As mentioned before, the MU takes these samples with a delay (measurement delay) which causes data shift error. This error causes uncertainty in the monitoring process of SAS and must be removed [12]. Fig. 1(a) shows a comparison between the amplitude of actual and measured data in a period of the TL signal (20ms). In Fig. 1(a), $\Delta t$ is the measurement delay that follows a random normal distribution function [19], and $\Delta A$ is data shift error caused by $\Delta t$. The amplitude of the TL signal can be equated as (1) [17]:

$$A(t) = A_{rms} \times \text{Cos}(2\Pi f t \pm \phi) \qquad (1)$$

where $A$ is the amplitude of actual data through TL, $A_{rms}$ denotes the root mean square of $A$, $f$ is the power system frequency, and $\phi$ refers to the power factor. As shown in Fig. 1(a), the measured data are sampled with a measurement delay ($\Delta t$) compared to actual data ($A$) and is equated as (2):

$$\begin{aligned} A_{measured}(t) &= A(t - \Delta t) \\ &= A_{rms} \times \text{Cos}(2\Pi f (t - \Delta t) \pm \phi) \end{aligned} \qquad (2)$$

where $\Delta t$ is the measurement delay value, $A_{measured}$ is the amplitude of measured data. Consequently, the data shift error ($\Delta A$) can be defined as the difference between $A_{measured}$ and $A$ ($\Delta A = A_{measured}(t) - A(t)$). Hence, we can calculate $A$ by having $\Delta t$ and $A_{measured}$, according to (3):

$$\begin{aligned} A(t) &= A_{measured}(t + \Delta t) \\ &= A_{measured}(t) - \Delta A \end{aligned} \qquad (3)$$

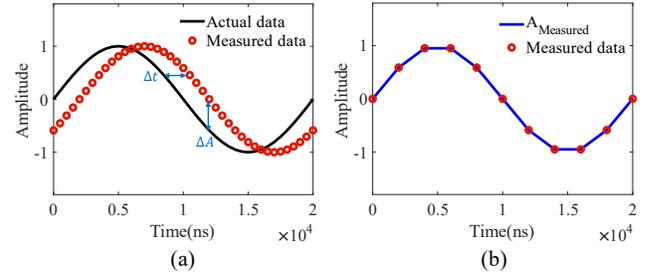

Fig. 1. (a) Amplitude of actual and measured data. (b) Reconstruction of $A_{measured}$ from discrete measured data.

According to (3), the process of removing $\Delta A$ from measured data is performed in two steps: 1) finding $\Delta t$ by calculating MU data sampling delay and 2) reconstructing the continuous form of $A_{measured}$ from discrete measured data and re-framing $A_{measured}$. The continuous form of $A_{measured}$ can be obtained by applying interpolation techniques to the discrete measured data, as illustrated in Fig. 1(b). One of the primary interpolation algorithms is the Lagrange interpolation algorithm [20]. The idea of Lagrange interpolation is the fact that by having n+1 pairs of $(x_i, y_i)$ of any function, it can be interpolated with a polynomial of the degree of n. The Lagrange interpolation algorithm supposes that for a given set $(x_0, y_0), (x_1, y_1), \ldots, (x_n, y_n)$, where no two $x_i$ are the same, the Lagrange polynomial is the polynomial of the least degree that maps each $x_i$ to the corresponding $y_i$. The Lagrange polynomial is equated as (4) where $x_i$ is i$^{th}$ sample point, $y_i$ is the function value corresponding to $x_i$ and $L(x)$ is the result of the Lagrange interpolation to the input x.

$$L(x) = \frac{(x-x_0)(x-x_2)...(x-x_n)}{(x_1-x_0)(x_1-x_2)...(x_1-x_n)} y_1 +$$
$$\frac{(x-x_0)(x-x_1)...(x-x_{i-1})(x-x_{i+1})...(x-x_n)}{(x_2-x_0)(x_2-x_1)...(x_i-x_{i-1})(x_i-x_{i+1})...(x_2-x_n)} y_2$$
$$+...+ \frac{(x-x_0)(x-x_2)...(x-x_n)}{(x_1-x_0)(x_n-x_2)...(x_n-x_n)} y_n \quad (4)$$

The linear compression of (4) is constructed as (5) [20] where n is the degree of interpolation and $\ell_i(x)$ is $i^{th}$ Lagrange interpolation coefficient corresponding to input x and is equated in (6).

$$L(x) := \sum_{i=0}^{n} y_i \ell_i(x) \quad (5)$$

$$\ell_i(x) := \prod_{\substack{0 \le m \le n \\ m \ne i}} \frac{x - x_m}{x_i - x_m} \quad (6)$$

### III. PROPOSED DATA FRAME CORRECTION SYSTEM

As mentioned before, the data shift error ($\Delta A$) can be removed by calculating measurement delay ($\Delta t$) and re-framing $A_{measured}$. Hence, to remove $\Delta A$ from the measured data, two primary steps have been taken in the proposed DFC system:
1. Calculating $\Delta t$ by finding the difference between MU sampling checkpoints (pre-defined time points at which MU must take samples from the TL signal [1]) and DRDY signal (DRDY signal is issued when MU samples the last measured data from each TL signal period [2]).
2. Using interpolation techniques to reconstruct and re-frame $A_{measured}$ in order to remove $\Delta A$ from the measured data.

Fig. 2 shows a general block diagram of our DFC system. In the first step, Measurement Delay Computation (MDC) Block calculates Measurement Delay ($\Delta t$) by finding the difference between the DRDY issued time and sampling checkpoint. After that, Measurement Delay is forwarded to Interpolation Block. In the second step, MU sends Measured Data to the Interpolation Block. Thereafter, the Interpolation Block uses the Lagrange interpolation algorithm to calculate Actual Data by re-framing Measured Data. The details of these two blocks are explained in the following subsections.

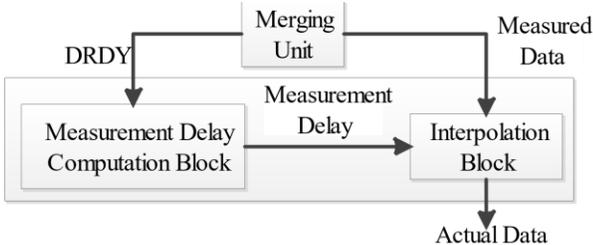

Fig. 2. Proposed DFC system.

#### A. Measurement Delay Computation (MDC) Block

Fig. 3 shows the MDC Block which calculates the Measurement Delay ($\Delta t$) in three steps:
1. Generating a pulse signal (Sampling Pulse (SP) in Fig. 3) with the period of MU sampling checkpoints.
2. Calculating the Measurement Delay by finding the difference of SP and DRDY signal.
3. Evaluating the accuracy of SP signal generation.

In the first step, the Sampling Checkpoint Generator (SCG) Block generates the SP signal with the period equals to that of of MU sampling checkpoints. Then, the SP signal is forwarded to both the Primary Counter and the Synchronization Blocks. In the second step, the Primary Counter Block counts the number of rising edges of the clock (Clk) between the SP signal edge rise time and the time the DRDY signal is issued. This number of Clk rising edges counted by the Primary Counter Block is Measurement Delay and is sent to the Interpolation Block afterward. In the third step, the global positioning system (GPS) module sends the one pulse per second (1PPS) signal to the Synchronization Block. After that, the Synchronization Block evaluates the accuracy of the SP signal generation by comparing it to the 1PPS signal. Thereafter, the Synchronization Block reports the SP signal generation accuracy status to SCADA via Sync Status Signal.

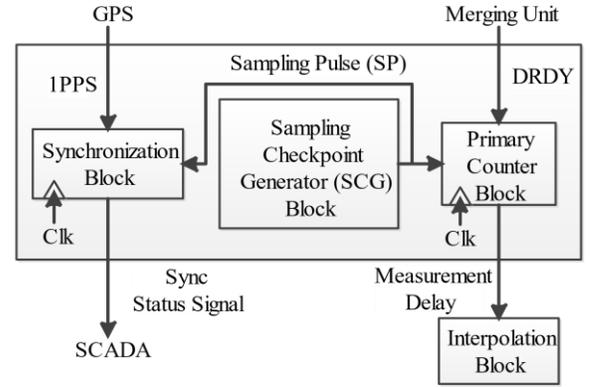

Fig. 3. Proposed MDC Block.

#### B. Interpolation Block

The Interpolation block removes the data shift error ($\Delta A$) from measured data in two steps:
1. Reconstructing $A_{measured}$ using Lagrange interpolation algorithm.
2. Removing data shift error by re-framing measured data.

In the first step, MU sends Measured Data to the Interpolation Block. After that, the Interpolation Block uses the Lagrange interpolation algorithm to reconstruct $A_{measured}$ from Measured Data frame based on (6). In the second step, MDC Block forwards the Measurement Delay to the Interpolation Block. Then, this Block calculates Actual Data by shifting $A_{measured}$ backwards in time-axis with the value of $\Delta t$ according to (3).

### IV. FPGA-BASED IMPLEMENTATION OF THE PROPOSED DATA FRAME CORRECTION SYSTEM

In this section, the FPGA implementation of the proposed DFC system is explained. It should be noted that all blocks are designed based on a clock with a 100MHz frequency (Spartan-

7 xc7s25ftgb196-2 internal clock [21]), every Reset input has an intrinsic delay, and the least significant bit (LSB) is B0.

*A. Measurement Delay Computation (MDC) Block*

Fig. 4 shows the proposed MDC Block architecture whose functionality is to calculate measurement delay ($\Delta t$) by comparing DRDY and SP signals. According to Fig. 4, the MDC Block computes the Measurement Delay in three steps. In step 1, SCG Counter generates Sampling Pulse (SP) based on the Clk pulse (100MHz) using SCG Counter and AND_1. Every time the SCG Counter counts 2,000,000 Clk rising edges (equal to 20ms), the AND_1 output (SP) is set to 1. After that, SP is forwarded to both the Synchronization Counter and the SR_Latch. Then, SCG Counter is reset by SP, and step 1 is finished. In step 2, the Q output of SR_Latch is set to 1 by SP; consequently, the Primary Counter is enabled and starts counting. The counting is continued until MU issues the DRDY signal. As the DRDY signal is issued, the output of OR_1 is set to 1 and the Q output of SR_Latch is set to 0; therefore, the Primary Counter is disabled and stops counting. Then, the output of the Primary Counter (Measurement Delay) is stored in Register_1, the Primary counter is reset, and step 2 is finished. It should be noted that the MDC Block must report the data loss (it happens when MU fails to send the data frame sample) to the SCADA. Thus, in step 2, if the Primary Counter counts beyond the sampling period of MU data frame ($20ms$), the Carry out of the Primary Counter (Data Lost Signal) is set to 1 and is stored in D_Flip_flop_2 to be sent to the SCADA. Thereafter, the OR_1 output is set to 1; consequently, the Q output of SR_Latch is set to 1, the Primary Counter is reset, and step 2 is finished.

The accuracy of SP generation must be evaluated every second [12] by comparing SP and the 1PPS signal of GPS. Hence, after 50 times of repeating step 1 (equal to $50 \times 20ms = 1s$), step 3 is begun. Each time step 1 is finished, the SP edge rises, and Synchronization Counter counts once. After 50 times of SP edge rising, if SP is generated accurately, the output of AND_2 must be 1; otherwise, a fault has happened in MDC Block, and it must be reported to SCADA via Sync Status Signal. Hence, each second, the output of XOR_1 (Sync Status Signal) is stored in D_Flip_flop_1 and is sent to SCADA.

*B. Interpolation Block*

From a hardware implementation point of view, the first challenge is that the Lagrange interpolation algorithm utilizes a massive amount of hardware resources [4]. To be more specific, the number of arithmetic units required for the hardware implementation of the Lagrange interpolation is equated as (7) where n is the degree of the interpolation algorithm and SUB_MUL_ADD_DIV is the summation of the number of subtractors, multipliers, divisions, and adders.

$$SUB\_MUL\_ADD\_DIV = 2\times(n-1) + n(n-1) + n(n-1) + \frac{n(n-1)}{2} + \frac{n(n-1)}{4} + ... + \frac{n(n-1)}{2} + \frac{n(n-1)}{4} + ... \approx 4n(n-1) \quad (7)$$

According to (7), for the Lagrange interpolation of the degree 256, the number of arithmetic units is 261630. These arithmetic units are 510 subtractors, 65280 dividers, 65280 adders, and 130560 multipliers, which require a massive amount of hardware resources to be implemented.

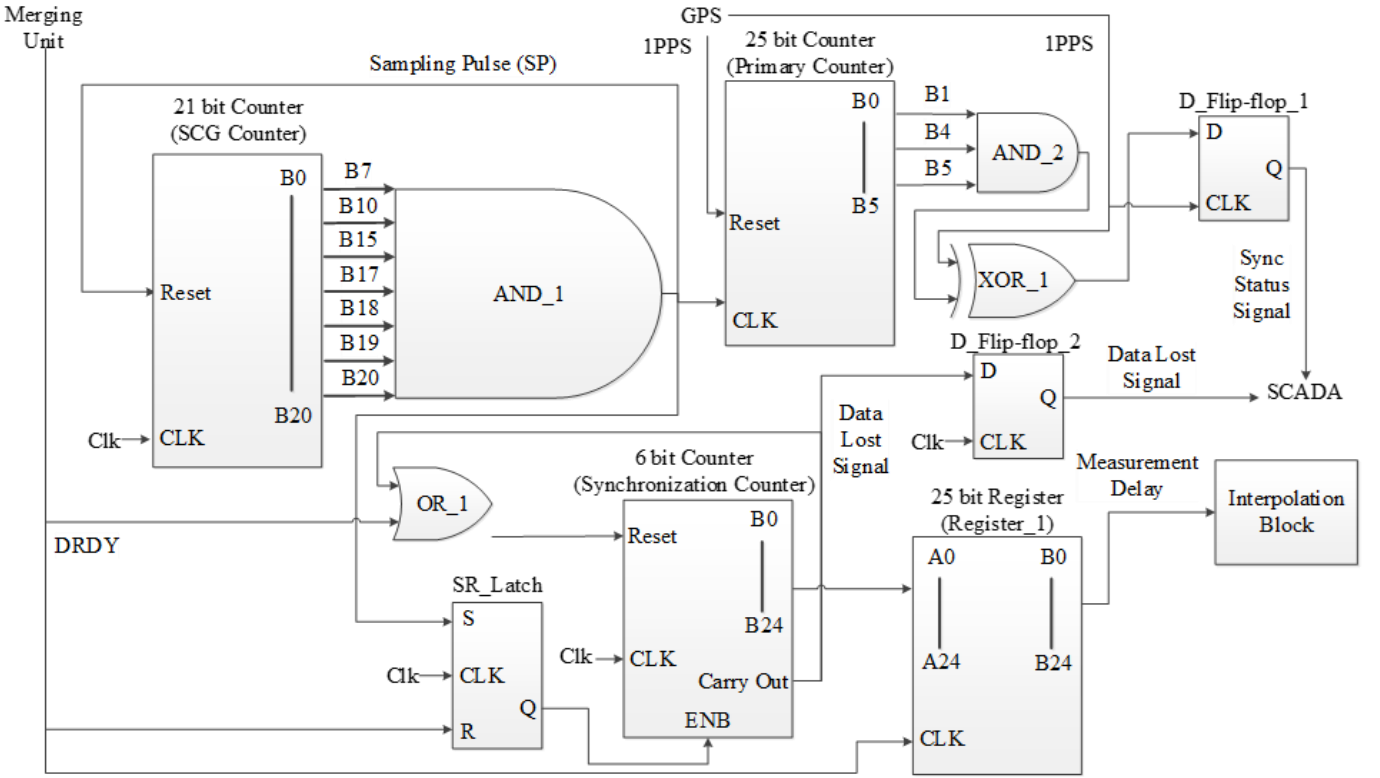

Fig. 4. Proposed MDC Block architecture.

The second challenge of the Lagrange interpolation hardware implementation is using a massive amount of memory. The memory cost of the Lagrange interpolation hardware implementation follows (8) where m is the register size of measured data time tag (m is the register size of $x_i$ in $(x_i, y_i)$), and 64 is the output memory bits requirement of floating-point calculation.

$$Memory\ (bit) = m \times n + 64 \times SUB\_MUL\_ADD\_DIV \quad (8)$$

As the $x_i$ is within 0 to 20,000,000ns range, to store each $x_i$, we need 25 bits of memory. Hence, for the Lagrange interpolation with the degree of 256, we need 16,750,720 bits of memory which is a massive memory requirement for the FPGA. To overcome these challenges, three approaches are presented which are described in the following paragraphs:
1. Analyzing the resolution requirement of DFC system to find the least degree of interpolation and fraction points that satisfies the accuracy requirements in IEC61850 [1].
2. Presenting a time tag mapping system to decrease the memory usage of storing measured data time tags
3. Utilizing a pre-processed weight matrix (W) to diminish division calculations in DFC system.

In the first approach, we elaborate on finding the least interpolation degree and fraction points that satisfy the accuracy requirement of the IEC61850 standard [1] based on the following analysis. In IEC61850 standard, the maximum acceptable error of measured data is $10^{-3}$ for Current and $25 \times 10^{-3}$ for Voltage measurements [18]. Hence, if our DFC system has a resolution of 0.001, we can guarantee that the DFC system will ultimately diminish the error of the shifted measured data. The resolution of the DFC system is based on two factors: 1) the system's interpolation degree and 2) the resolution of system calculation fractions.

The purpose of using the Lagrange interpolation algorithm is to reconstruct the continuous function of $A_{measured}$ from discrete measured data. MU samples all these measured data within the 0-20ms range in each period of TL signal. As the clock period of the DFC system is 10ns, the resolution of measurement delay ($\Delta t$) calculation is 10ns. Thus, $\Delta t$ of all measured data is a multiple of 10ns. Therefore, the minimum interpolation degree is the least degree in which the Lagrange interpolation algorithm can anticipate $A_{measured}(t)$ (where t is a multiple of 10ns within 0-20ms) with the maximum absolute error less than 0.001. Fig. 5 shows the flow chart for the determination of the minimum interpolation degree. To determine the minimum interpolation degree (N), we took 256 (the number of data in each frame) samples $(x_i, y_i)$ of a sinusoidal signal where $x_i$ is across 0-360° with a constant step of 1.4° and $y_i = \sin(x_i)$. Thereafter, Lagrange interpolation algorithm with different interpolation degrees ($N \geq 3$) is applied to $(x_i, y_i)$ to reconstruct $A_{measured}(t)$. Then, for each N≥ 3, we calculated all possible measured data which resulted in 2,000,000 samples of $A_{measured}(t)$. After that, we computed the absolute error of every 2,000,000 samples of $A_{measured}(t)$ for each N and chose the least N with the maximum absolute error less than 0.001 as the minimum degree of interpolation. Fig. 6 demonstrates the maximum absolute error of a total of 2,000,000 samples of $A_{measured}(t)$ for different interpolation degrees (N). According to this figure, the minimum value of N to satisfy the accuracy requirement of the IEC61850 standard [1] (maximum absolute error < 0.001) is 16 in the floating-point calculation system.

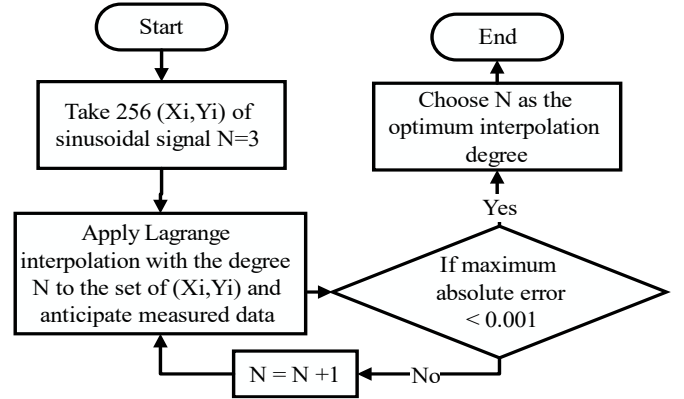

Fig. 5. The flow chart for determination of the minimum interpolation degree.

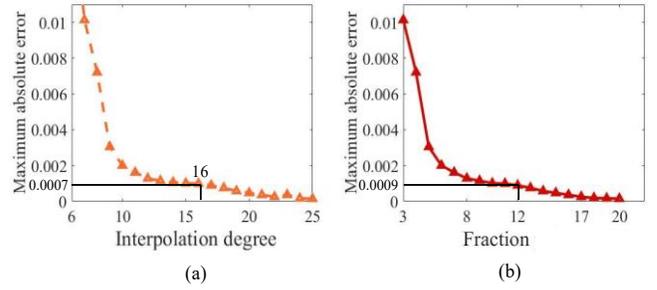

Fig. 6. (a) Maximum absolute error versus interpolation degree. (b) Maximum absolute error versus fraction points.

To avoid hardware implementation elaboration caused by using floating-point calculation system, we investigated the possibility of the hardware implementation of the Lagrange interpolation via fixed-point. To evaluate the fixed-point system option, we apply the Lagrange interpolation with N=16 to the discrete measured data and reconstructed continuous function of $A_{measured}$ 32 times. Each time, we kept the different number of fractions (0-32) of calculation results. Then, for each number of fractions, we calculated all possible measured data (2,000,000 samples of $A_{measured}(t)$ as mentioned in the previous paragraph) and found the maximum absolute error of these samples. Thereafter, we chose the least number of fractions with the maximum absolute error less than 0.001 as the minimum number of fractions. Fig. 7 shows the maximum absolute error of a total of 2,000,000 samples of $A_{measured}(t)$ with keeping a different number of fractions. These results show that keeping 12-bit fractions of calculations is adequate for obtaining the accuracy of 0.001. Hence, to implement the Lagrange interpolation, which satisfies the accuracy requirement of IEC61850 standard [1] (maximum absolute error < 0.001), we chose N=16 instead of N=256 and kept 12-bits fractions of calculations instead of 64-

bits floating-point.

In the second approach, we present a time tag mapping system to reduce the memory cost of storing data. The time tag of measured data ($x_i$ in $(x_i, y_i)$) is from 0 to 20,000,000ns [1], which requires 25 bits to be stored. Each data frame contains 256 data [19], storing data frames necessities allocating $256 \times 25 = 6400$ bits of memory. To decrease the memory usage, we map the time tags from [0:20,000,000] to [0:256] steps where every 78125ns (data sampling period) considered one step. This mapping system decreases the register size of time tags from 25 to 8 bits. The mapping formula is according to (9).

$$x_i(step) = \frac{x_i(ns)}{78125} = x_i(ns) \times 0.000128 \quad (9)$$

Performing division in digital system cause system elaboration and accuracy reduction; therefore, it should be avoided. Hence, as the third approach, we utilize a pre-processed weight matrix to diminish division calculations in DFC system. As the MU takes a sample of each measured data every 78125ns, $x_i$ and $x_m$ in the calculation of Lagrange interpolation coefficient (according to (5)) are pre-defined values [1]. Hence, it is plausible to calculate a pre-processed weight matrix ($W$) that includes results of $\ell_i$ divisions according to (10). This allows us to perform interpolation calculations only by summation and multiplications, avoiding divisions. Therefore, $\ell_i(x)$ can be calculated without any divisions according to (11).

$$W_{i,m} = \frac{1}{(x_i - x_m)} \quad (10)$$

$$\ell_i(x) = \prod_{\substack{0 \prec m \prec n \\ m \neq i}} (x - x_m) W_{i,m} \quad (11)$$

In the following paragraphs, we will explain the hardware implementation of proposed the proposed Interpolation Block using three above mentioned approaches. The following equation can describe the functionality of the proposed Interpolation Block for each measured data:

$$A(x) = A_{measured}(x + \Delta t)$$
$$= \sum_{i=0}^{16} y_i \ell_i(x + \Delta t) \quad (12)$$

where $A(x)$ is the actual value of TL signal in time (x), $\Delta t$ is measurement delay, $A_{measured}(x + \Delta t)$ is the measured data in x, $y_i$ is ith measured data in each data frame, and $\ell_i$ is ith Lagrange interpolation coefficient of $A_{measured}$. Based on (12), each A(x) requires 15 $\ell_i(x)$ to be calculated in the Interpolation block. As all 15 $\ell_i(x)$ do not have any data dependency with each other; it is possible to calculate all of them simultaneously. Fig. 7 shows the proposed Interpolation Block architecture. As shown in in Fig. 7, the proposed Interpolation Block computes A(x) in three steps. In step 1, the Measurement Delay is sent to all Lagrange Coefficient Computation Blocks. In the same time, the required division results of each Lagrange Coefficient Computation Block ($W$) are forwarded to Lagrange Coefficient Computation Blocks. Thereafter, Lagrange Interpolation Coefficient Computation Blocks calculate $\ell_i(x)$ (where i is equal to the number of Lagrange Coefficient Computation Block) and step 1 is finished. In step 2, Memory Divider stores measured data from MU and sends $i^{th}$ Measured Data to the related Multiplier_i. Then, each $\ell_i(x)$ is multiplied by $i^{th}$ Measured Data. After that, the results of 15 Multipliers are 15 required $\ell_i(x)$ for calculating A(x). Hence, in step 3, the Adder Block calculates the summation of all outputs of Multiplier_i (i=1, 2, …, 15). The output of the Adder Block.

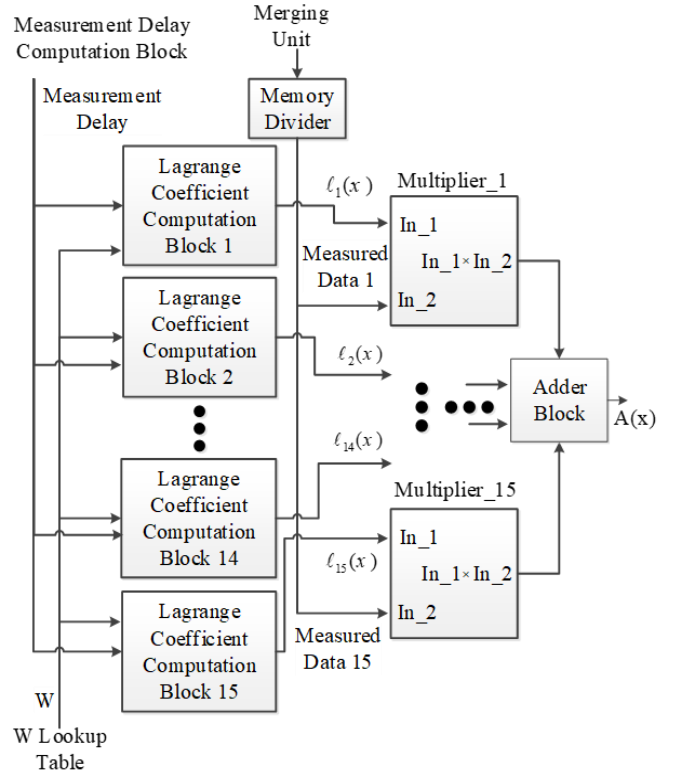

Fig. 7. Proposed Interpolation Block architecture.

The functionality of Lagrange Coefficient Computation Block is according to (13):

$$\ell_j(x + \Delta t) = \prod_{\substack{0 \prec m \prec 16 \\ m \neq i}} Z_{i,m} \quad (13)$$

where $Z_{i,m}$ is the $m^{th}$ Lagrange interpolation sub-coefficient of $i^{th}$ Lagrange interpolation coefficient corresponding to input x and is equated as (14).

$$Z_{i,m} = ((x + \Delta t) - x_m) W_{i,m} \quad (14)$$

Same as $\ell_i(x)$, the $Z_{i,m}$ are independent of each other. Thus, it is possible to calculate all of them simultaneously. Fig. 8 shows the Lagrange Coefficient Computation Block architecture. According to Fig. 8, each Lagrange Coefficient Computation Block computes $\ell_i(x)$ in two steps. In step 1, Measurement Delay and related divisions results of each Lagrange Sub-coefficient Computation Block ($W_{i,m}$ where m is the number of Lagrange Sub-coefficient Computation Block, and i equals the number of Lagrange Coefficient Computation Block in which the Lagrange Sub-coefficient Computation Block is) are forwarded to Lagrange Sub-coefficient Computation Blocks. Then, Lagrange Sub-coefficient Computation Blocks compute all $Z_{i,m}$ at the same time and forward them to First_Layer_Multipliers. In step 2, all $Z_{i,m}$ are multiplied by each other's according to (13) in four layers of series multipliers. The output of the Forth_Layer_Multiplier is $\ell_i(x)$ The Lagrange Sub-coefficient Computation Block architecture is according to Fig. 9 and its detail is explained in the following paragraph.

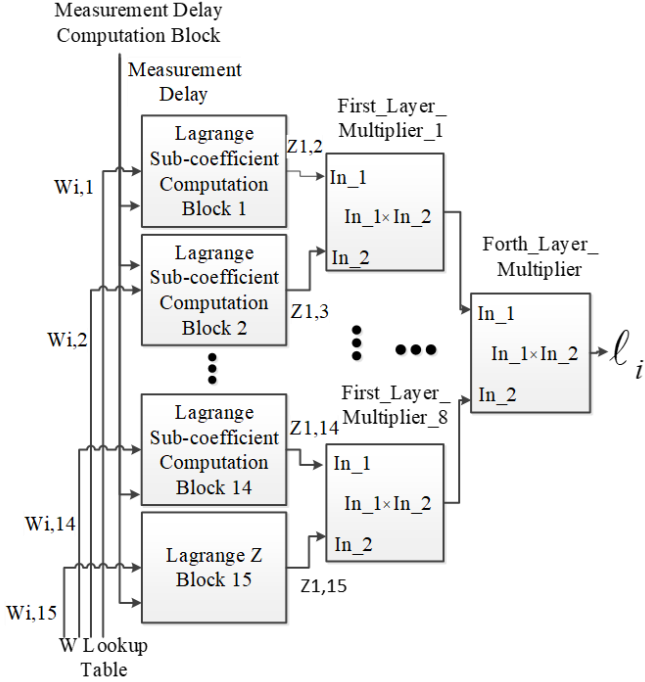

Fig. 8. Lagrange Coefficient Computation Block architecture.

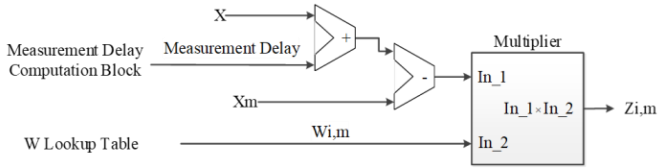

Fig. 9. Lagrange Sub-coefficient Computation Block architecture.

Each Lagrange Sub-coefficient Computation Block functionality is to compute a $Z_{i,m}$. As shown in Fig. 9, the Lagrange Sub-coefficient Computation Block consists of three blocks: 1) an adder block, 2) a subtractor block, and 3) a Multiplier block. The adder block computes $x + \Delta t$ and forwards the results of summation to the subtractor block. The subtractor block calculates $(x + \Delta t) - x_m$ (where m equals the number of Lagrange Sub-coefficient Computation Block) and its result is sent to the Multiplier block. Finally, in the Multiplier block, to calculate the $Z_{i,m}$ (where m is the number of Lagrange Sub-coefficient Computation Block and i equals to the number of Lagrange Coefficient Computation Block in which the Lagrange Sub-coefficient Computation Block is), the result of the subtractor block is multiplied by $W_{i,m}$ according to (14).

V. EXPERIMENTAL RESULTS AND DISCUSSION

In this section, we made a quantitative comparison between accuracy, processing time, and utilization cost of the available and proposed DFC systems for the MUs. The authors of [4] and [9] did not report the time and hardware resources cost of their implementation. Hence, we implemented their DFC system on the Spartan-7 xc7s25ftgb196-2, the same hardware we implemented our proposed DFC system, to make comparisons.

To test the accuracy of the proposed DFC system, we have corrected a frame of 256 measured data from the period of a sinusoidal signal frame, which has a random Gaussian data shift and a partial discharge (PD) [22] according to Fig. 10(a). In Fig. 10(a), the shown corners are the points with the maximum interpolation errors (as they are extremum points [23]). Hence, all DFC systems results should satisfy the accuracy requirement of IEC61850 [1] in these points. The absolute error of existing and proposed DFC system in these points are according to Fig. 10(b). According to Fig. 10(b), the proposed DFC system satisfies the accuracy requirement of IEC61850 standard [1] because the maximum absolute error of corrections of all corners are less than 0.001. Table I includes the maximum absolute error and resolution of existing and the proposed DFC system.

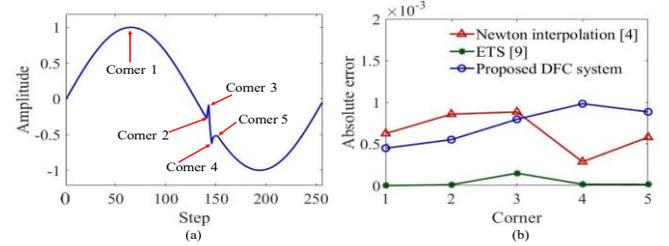

Fig. 10. (a) Corners of the sinusoidal signal with a random gaussian data shift error. (b) Absolute errors in corners.

TABLE I
THE COMPARISON OF DIFFERENT ARCHITECTURES AND PROPOSED ARCHITECTURE ACCURACY

| DFC System | Maximum Absolute Error of Corrections in Corners | Resolution |
|---|---|---|
| [4] | 0.00023 | 0.000184 |
| [9] | $4.71 \times 10^{-14}$ | $1.64 \times 10^{-14}$ |
| **Proposed System** | **0.00098** | **0.000244** |

To compare the processing time of each DFC system, a set of 256 shifted data is given to each DFC system to be corrected. The DFC system in [9], which uses a polynomial interpolation with the degree 104, requires 76.3ms to correct the frame of 256 data. The DFC system in [4] corrects the data frame in 13.1ms using the Newton interpolation algorithm with the interpolation degree of 25. The proposed DFC system, however, corrects the data frame in 2.1ms. The Lagrange interpolation algorithm is not recursive compared to the Newton interpolation; therefore, it provides the possibility of parallel interpolation calculation. Moreover, reducing the interpolation degree from 256 to 16 and utilizing the W matrix to remove calculation divisions decrease the proposed DFC system processing time 83.9% and 97.2% compared to the [4] and [9], respectively. Table II compares the processing time and interpolation algorithm of existing and the proposed DFC system. According to Table II, the DFC system in [4] and the proposed DFC system can achieve data frame correction in a standard time budget (20ms in [2]). Hence, in real-time monitoring applications, [4] and the proposed DFC system are satisfying the time budget.

TABLE II

THE PROCESSING TIME OF STATE-OF-ART AND PROPOSED ARCHITECTURE

| DFC System | Interpolation Algorithm | Interpolation Degree | Processing Time (ms) |
|---|---|---|---|
| [4] | Newton | 25 | 13.1 |
| [9] | Linear Polynomial | 104 | 76.3 |
| **Proposed System** | **Lagrange** | **16** | **2.1** |
| Frequency | | 100 MHz | |

Table III shows a comparison of the hardware resource utilization of existing and the proposed DFC system. As shown in Table III, the Proposed DFC system utilizes 50% and 85.7% less BRAM, 98% and 99.8% lees DSP, 96.2% and 99.6% less FF, and 95.7% and 99.6% less LUT compared to [4] and [9] respectively. It is worth mentioning that all designs are implemented on the Spartan-7 xc7s25ftgb196-2. The employment of input time tagging mapping system and using fixed-point calculation system significantly decrease proposed DFC system hardware resource utilization.

TABLE III

THE RESOURCE UTILIZATION OF STATE-OF-ART AND PROPOSED ARCHITECTURE

| | | BRAM | DSP | FF | LUT |
|---|---|---|---|---|---|
| [4] Newton DFC System | Total | 2 | 1048 | 70580 | 40287 |
| | Utilization | ~0% | 7% | 2% | 3% |
| [9] ETS DFC System | Total | 7 | 18765 | 643401 | 517811 |
| | Utilization | ~0% | 13% | 5% | 11% |
| Proposed DFC System | Total | 1 | 20 | 2662 | 1720 |
| | Utilization | ~0% | 3% | 6% | 2% |
| Available | | 45 | 80 | 29200 | 14600 |

VI. CONCLUSION

In this paper, we proposed a data frame correction (DFC) system based on the measured data phase. The proposed DFC system uses the difference time of data frame sampling of MU and SAS sampling checkpoints to remove data shift error without using a reference signal. The proposed DFC system is implemented as a hardware accelerator on FPGA. Compared to the amplitude-based DFC systems, the results show that the proposed DFC system can achieve data correction up to 99.6% less hardware resources utilization and fulfills 9× calculation speed while maintaining the IEC61850 required accuracy in 2.1ms. Hence, it can be used in real-time monitoring applications.

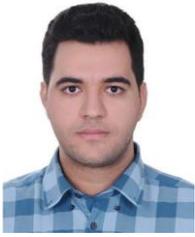 **Mohammad Hashemi** was born in Ahvaz, Iran, in 1994. He received the B.Sc. degree in electrical engineering from Islamic Azad University, Tehran, Iran, in 2017. Since 2018, he has been pursuing the M.Sc. degree in electrical engineering at University of Tehran, Tehran, Iran. His current research interests include embedded system design, FPGA-based hardware accelerators design, machine learning and its applications.

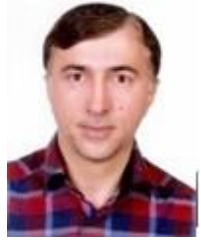 **Bijan Alizadeh** (SM'13) received his Ph.D. degree in electrical and computer engineering from the University of Tehran, Iran, in 2004. He was with the School of Electrical Engineering, Sharif University of Technology, Iran, as an Assistant Professor from 2005 to 2007 and VDEC, The University of Tokyo, Japan, as a Research Associate from 2007 to 2010. He has been Assistant Professor with the School of Electrical and Computer Engineering at the University of Tehran since 2011, where he is currently an Associate Professor. He has authored or co-authored over 130 publications in international scientific journals and conferences. He has been engaged in the research and development of VLSI systems, FPGA-based reconfigurable computing, hardware Trojan detection, IoT authentication mechanisms, formal verification and debug.